\documentclass[conference]{IEEEtran}

\usepackage{cite}
\usepackage{bm}
\usepackage{amsmath}
\usepackage{mathrsfs}

\usepackage{algorithmic}

\usepackage{array}

\ifCLASSOPTIONcompsoc
\usepackage[caption=false,font=normalsize,labelfont=sf,textfont=sf]{subfig}
\else
\usepackage[caption=false,font=footnotesize]{subfig}
\fi
\usepackage{CJK}
\usepackage{url}
\usepackage{graphicx,amsmath,amssymb,amsfonts}
\usepackage{algorithmic,algorithm}

\usepackage{xcolor}

\newtheorem{theorem}{\textbf{Theorem}}

\newtheorem{lemma}{\textbf{Lemma}}

\newtheorem{corollary}{\textbf{Corollary}}

\makeatletter

\newcommand{\Rmnum}[1]{\expandafter\@slowromancap\romannumeral #1@}
\makeatother

\begin{document}
\bstctlcite{ref:BSTcontrol}

\title{Deep Reinforcement Learning for Energy-Efficient Beamforming Design in Cell-Free Networks}


\author{
	\IEEEauthorblockN{ 
		Weilai~Li, 
		Wanli~Ni, 
		Hui~Tian,
		and~Meihui~Hua
		\IEEEauthorblockA{
			State Key Lab. of Networking and Switching Technology, Beijing Univ. of Posts and Telecommun., Beijing, China\\
			E-mail: \{liweilai, charleswall, tianhuin, huameihui\}@bupt.edu.cn
		}
	}
}

\maketitle

\begin{abstract}
Cell-free network is considered as a promising architecture for satisfying more demands of future wireless networks, where distributed access points coordinate with an edge cloud processor to jointly provide service to a smaller number of user equipments in a compact area. In this paper, the problem of uplink beamforming design is investigated for maximizing the long-term energy efficiency (EE) with the aid of deep reinforcement learning (DRL) in the cell-free network. Firstly, based on the minimum mean square error channel estimation and exploiting successive interference cancellation for signal detection, the expression of signal to interference plus noise ratio (SINR) is derived. Secondly, according to the formulation of SINR, we define the  long-term EE, which is a function of beamforming matrix. Thirdly, to address the dynamic beamforming design with continuous state and action space, a DRL-enabled beamforming design is proposed based on deep deterministic policy gradient (DDPG) algorithm by taking the advantage of its double-network architecture. Finally, the results of simulation indicate that the DDPG-based beamforming design is capable of converging to the optimal EE performance. Furthermore, the influence of hyper-parameters on the EE performance of the DDPG-based beamforming design is investigated, and it is demonstrated that an appropriate discount factor and hidden layers size can facilitate the EE performance.
\end{abstract}

\section{Introduction}
An innovative network architecture called cell-free network emerges with myriad of attention recently, which is considered to bring numerous potentials for the future wireless networks \cite{ref02,ref03,Max}.\let\thefootnote\relax\footnotetext{This paper is funded by Beijing Univ. of Posts and Telecommun.-China Mobile Research Institute Joint Innovation Center.} There are no cells or cell boundaries in the cell-free network. All access points (APs) are fully connected to a smaller number of user equipments (UEs) and coordinate together with an edge cloud processor (ECP) in a compact area, in order to promote the exploitation of favorable propagation and channel hardening, as well as mitigate the cell-edge problem \cite{ref05}. The APs estimate the channel locally based on the non-orthogonal pilots transmitted from UEs due to the insufficient orthogonal resources, leading in turn to pilot contamination \cite{jour02}. In order to alleviate inter-user interference, the non-orthogonal pilot contamination and excessive control signaling, one of the essential signal processing technologies, beamforming, is utilized in cell-free networks, and it is proved that appropriate beamforming design is conducive to improve the system throughput, energy efficiency (EE) and probability of coverage \cite{jour01,hua1,ref07}.

The state-of-the-art literature on beamforming design in cell-free networks fix attention on diverse aspects \cite{bf1,bf3,bf_EE,ref07,ref08}. In \cite{bf1}, the authors proposed an iterative algorithm by utilizing the max-min beamformer and derived the capacity lower bound of the cell-free network with channel estimation error. In \cite{bf3}, the authors presented a modification of conjugate beamforming for the forward link in cell-free networks that eliminated the self-interference completely and with no action required at the receivers. In \cite{bf_EE}, the authors applied a distributed conjugate beamforming scheme at each AP for the use of local channel state information (CSI) and the scheme could enhance the total EE.
However, the beamforming design is a problem of dynamic, successive decision-making under uncertainty, and these conventional optimization schemes are limited by their myopic decision criteria, poor scalability and high complexity. In view of it, deep reinforcement learning (DRL) is an adaptive method to overcome these challenges \cite{ML}. In \cite{ref07}, the authors formulated a novel hybrid model using deep deterministic policy gradient (DDPG) and deep double Q-network for jointly optimizing the clustering of APs and the beamforming vectors, the simulation results demonstrated that the model is efficient to maximize the per-user transmission rate. In \cite{ref08}, the authors proposed a distributed dynamic down-link-beamforming coordination method with partial observability of the CSI using Deep Q-Learning, and the simulation results proved that the method is effective for the improvement of the achievable transmission rate. The literature above reveal the effectiveness of DRL in beamforming optimization for network throughput improvement. Whereas, the energy consumption becomes a critical issue as the future network size scales up, and the trade-off between the throughput and the energy consumption deserves comprehensive attention in more practical scenarios.

Motivated by the aforementioned discussion, the main contributions of this paper are as follow: 
1) The expression of signal to interference plus noise ratio (SINR) for each UE is derived under minimum mean square error (MMSE) channel estimation and successive interference cancellation (SIC) detection.
2) We further define the closed-form expression of the long-term EE, and the DDPG algorithm is utilized for beamforming design to achieve the long-term EE maximum.
3) The convergence and optimality of our proposed DDPG-based beamforming design is analyzed theoretically and the EE performance is evaluated numerically, and its superiority in comparison with benchmarks is demonstrated. 
4) The influence of the hyper-parameters on the EE performance is further explored, and appropriate discount factor and hidden layers size improving the EE performance is concluded.


\section{System Model and Problem Formulation}
\subsection{System Model}
As illustrated in Fig. \ref{1}, a cell-free network with $M$ single-antenna APs and $K$ single-antenna UEs is considered, where the locations of APs are fixed, and UEs are initially randomly distributed which are assumed to be of low mobility. In the cell-free network, all APs are fully connected to all UEs and link to ECP via the perfect backhaul links. All UEs are uniformly served by the distributed APs in a collaborated manner. The channel estimation is conducted at each AP locally, and signal detection occurs in the  centralized processing unit (CPU) pool in ECP based on the channel state information (CSI) sent from APs. Subsequently, the CPU performs the beamforming design and returns the beamforming decision to all APs as feedback.

To acquire CSI, a random set of pilot sequences is assigned to UEs for channel estimation. Firstly, we assume that the channel between the k-th UE and the m-th AP is modeled as follow:
\begin{eqnarray}
\label{1}
{{g_{mk}} = \beta _{mk}^{1/2}{h_{mk}},
}	
\end{eqnarray}
	where $h_{mk}$ is the small-scale fading coefficient between the $k$-th UE and the $m$-th AP, and $\beta_{mk}$ is the large-scale fading coefficient, which is given by:
\begin{eqnarray}
\label{2}
{\beta _{mk}} = {\varsigma _0}{({d_{mk}})^{ - 2}},
\end{eqnarray}
where $\varsigma_0 =  - 30\rm{dB}$ is the path loss at the reference distance of 1 meter, $d_{mk}$ denotes the access link distance. We assume that ${h_{mk}}(\forall m,\forall k)$ are independent and identically distributed (i.i.d.) random variables, i.e., ${h_{mk}}\sim {\cal C}{\cal N}(0,1)$.

\subsection{Channel Estimation}
The method of the channel estimation is to assign pilot sequences to UEs in the coverage area. The pilot sequence of the k-th UE can be presented as ${\bm{\varphi}_k} = {\left[ {{\bm{\varphi}_{k,1}} \cdots {\bm{\varphi}_{k,{\tau_l}}}} \right]^H}$, $\| {\bm{\varphi} _k} \|^2 = 1$, where ${\tau_l}$ is less than the coherence time of the channel ${\tau_c}$. It’s worth mentioning that different UEs may be assigned the same pilot sequence on account of the limited pilot length ${\tau_l}$, i.e., ${\tau _l} \le K$, hence the pilot sequences are partially non-orthogonal which satisfies $\left| {\bm{\varphi}_k^H{\bm{\varphi} _n}} \right| \ne 0,k \ne n$.  In this way, the pilot signal received at the $m$-th AP can be expressed as:
\begin{eqnarray}
\label{3}
{\bm{y}_{m,p}} = \sum\limits_{k = 1}^K {\sqrt {{\tau _l}{\delta _l}} {g_{mk}}} {\bm{\varphi}_k} + {\bm{\sigma}_{m,p}},
\end{eqnarray}
where ${\delta _l}$ is normalized transmission power for each symbol of the $k$-th UE’s pilot vector, ${\bm{\sigma} _{m,p}} \in {{\mathbb{C}}^{{\tau _l} \times 1}}$ is the complex-valued additive white Gaussian noise (AWGN) vector related to pilot symbols with i.i.d. random variables, i.e., ${\bm{\sigma} _{m,p}}\sim {\cal C}{\cal N}(0,{\sigma ^{\rm{2}}})$.
\begin{figure}[t]
	\centering
	\includegraphics[width=3.5 in]{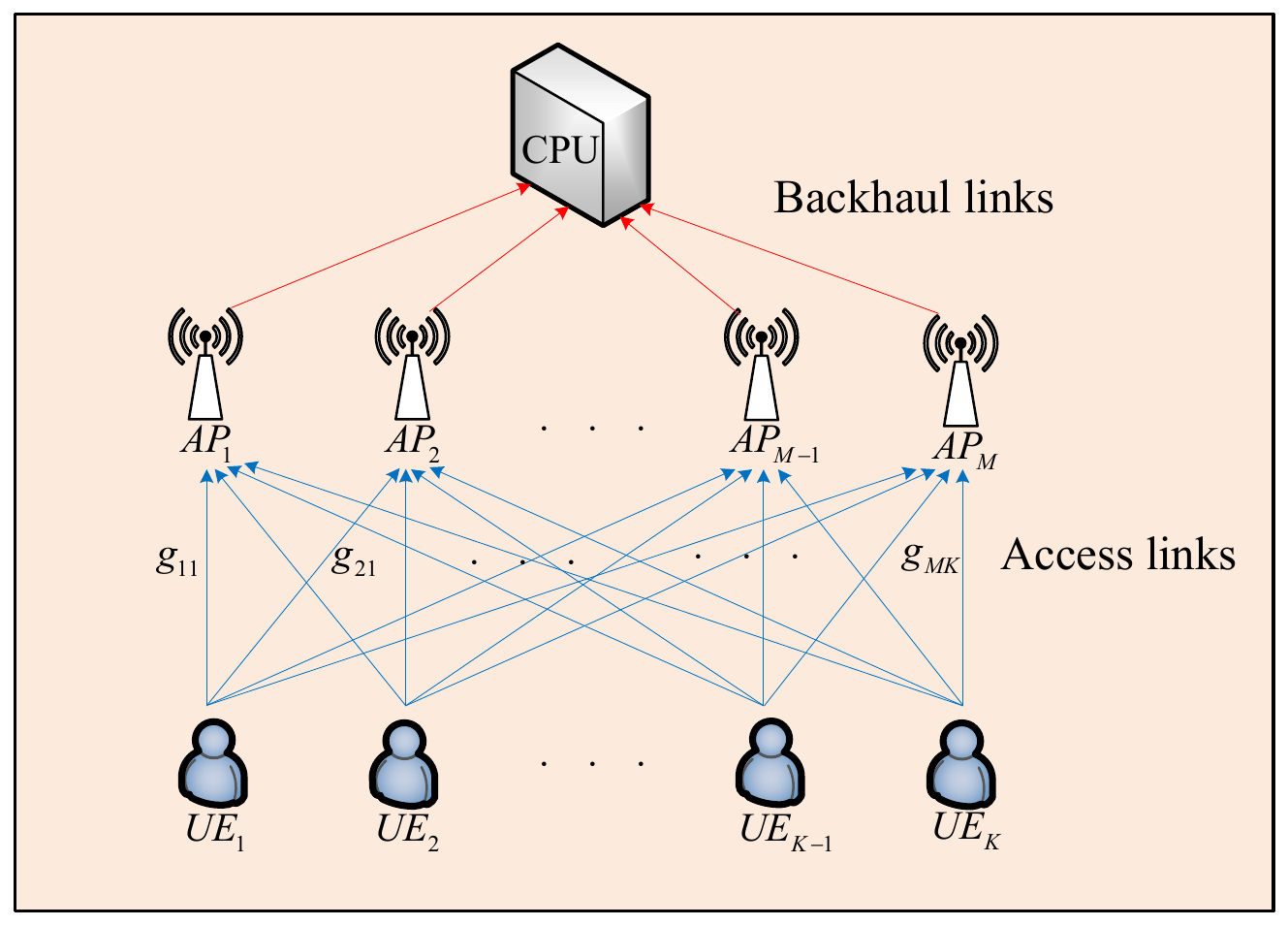}
	\caption{System model of cell-free networks.}
	\label{1}
\end{figure}

It is assumed that the large-scale fading coefficient $\beta_{mk}$ is known, while the small-scale fading coefficient ${h_{mk}}$ is unknown. In other words, the aim to estimate CSI is equivalent to estimate channel coefficient ${\hat g_{mk}}$. In the first phase, we denote ${\hat y_{mk}}$ as the projection of ${\bm{y}_{m,p}}$ onto $\bm{\varphi} _k^H$ and we can obtain:
\begin{eqnarray}
\label{7}
\begin{split}
\hat{y}_{mk}&={{\bm{\varphi}_k^H}}\bm{y}_{m,p} \\
&=\sqrt{\tau_l\delta_{l}}g_{mk}+\sqrt{\tau_l\delta_{l}}\sum\limits_{k'\ne k}^K g_{mk'}\bm{\varphi}_k^H\bm{\varphi}_{k'}+\bm{\varphi}_k^H\bm{\sigma}_{m,p}.
\end{split}
\end{eqnarray}

When we set the estimation coefficient ${\mu_{mk}}$ and ${\hat g_{mk}} = {\mu _{mk}} \cdot {\hat y_{mk}}$, the estimation error is $e = {\hat g_{mk}} - {g_{mk}}$ accordingly. By utilizing the MMSE criterion to minimize $\mathbb{E}\{{e^*}  e\}$ \cite{ref01}, we acquire the following equation:
\begin{eqnarray}
{{\partial {\mathbb{E}}\{{e^*}  e\}} \over {\partial {g_{mk}}}} = 0,
\end{eqnarray}
after a series of operations, the estimation coefficient ${\mu _{mk}}$ is calculated by:
\begin{eqnarray}
\label{6}
\begin{split}
&{\mu _{mk}} = {{{\mathbb{E}} [{\hat y^*_{mk}{g_{mk}}}] } \over {{\mathbb{E}}[ {{{\left| {{{\hat y}_{mk}}} \right|}^2}} ]}} \\
&\qquad\!= {{\sqrt {{\tau _l}{\delta _l}} {\beta _{mk}}} \over {{\tau _l}{\delta _l}\sum\nolimits_{n = 1}^K {{\beta _{mk}}{{\left| {\bm{\varphi} _k^H{\bm{\varphi} _n}} \right|}^2} + {\sigma} ^{\rm{2}}} }}.
\end{split}
\end{eqnarray}
Ultimately, ${\hat g_{mk}}$ can be formulated as ${\hat g_{mk}} = {\mu _{mk}}\bm{\varphi} _k^H{\bm{y}_{m,p}}$.

\subsection{Uplink Data Transmission}
In the cell-free network, the signal from each UE will be received by all APs. The APs weight the baseband signals by beamforming vector ${w_{mk}}$ and transmit the signals to the CPU pool through backhaul links. Intuitively, maximizing the desired signal and minimizing the interference, pilot contamination and noise conduces to improve the user experience. Signal of each UE is detected in the CPU pool, and the detected signal of the $k$-th UE can be expressed as:

\begin{eqnarray}
\label{7}
\begin{split}
& {y_k} = \sum\limits_{m = 1}^M {\sum\limits_{n = 1}^K {{w_{mk}}({{\hat g}_{mn}}\sqrt {{p_u}} {x_n} + {{\tilde{\sigma} }_m})} }   \\
& {\rm{  \quad  }} =  \sum\limits_{m = 1}^M {{w_{mk}}} \sum\limits_{n=1}^K {\sqrt {{\tau _l}{\delta _l}{p_u}} {x_n}} {\mu _{mn}}{g_{mn}} \\
& {\rm{ \quad  }} + \sum\limits_{m = 1}^M {{w_{mk}}} ( \sum\limits_{p=1}^K {\sum\limits_{q \ne p}^K {\sqrt {{\tau _l}{\delta _l}{p_u}} {x_p}} } {\mu _{mp}}\left| {\bm{\varphi} _p^H{\bm{\varphi }_q}} \right|{g_{mq}}) \\
& {\rm{ \quad  }} + \sum\limits_{m = 1}^M {{w_{mk}}} (\sum\limits_{s = 1}^K {\sqrt {{p_u}} {\mu _{ms}}{x_s}\left| {\bm{\varphi} _s^H{\bm{\sigma} _{m,p}}} \right| + } {\tilde \sigma _m}),
\end{split}
\end{eqnarray}
where $0 \le {w_{mk}} \le 1$ is the beamforming vector between the $k$-th UE and the $m$-th AP, ${p_u}$ is the UE’s uplink signal transmission power with ${\rm{0}} \le {p_u} \le {P_u}$, where ${P_u}$ is the maximum allowable signal transmission power, ${x_n}$ is the n-th UE’s transmitted symbol that satisfies ${\mathbb{E}}\{  |x_n|^2 \} = 1$, $\tilde \sigma _m$ is the AWGN at the the $m$-th AP with ${\tilde{\sigma} _m}\sim {\cal C}{\cal N}(0,{\sigma ^{\rm{2}}})$.

The equation (7) is composited of three components: the first component is the desired signal mixed with inter-user interference, the second component is non-orthogonal pilot contamination, and the last component is AWGN-related estimation error and AWGN component. For the purpose of the enhancement of the SINR, first, we can simplify the elements in (7):
\begin{subequations}\label{8}
	\begin{eqnarray}
	\label{6a}
	&\widetilde{g}_{mn}=\sqrt{\tau_l\delta_{l}p_u}\mu_{mn}g_{mn},\\ 
	&\widetilde{g}_{mq}=\sqrt{\tau_l\delta_{l}p_u}\mu_{mp}|\bm{\varphi}_p^H\bm{\varphi}_q|\,g_{mq}, \\
	&\psi_k=\sum_{s=1}^Kp_u\mu_{ms}^2\bm{\varphi}_s^2\bm{\sigma}_{m,p}^2+\tilde{\sigma}_m^2.
	\end{eqnarray}
\end{subequations}
Accordingly, we derive the closed-form expression of the $k$-th UE's SINR from (7):
\begin{eqnarray}
\label{7}
\begin{split} 
{\gamma_k} =\dfrac{\sum\limits_{m = 1}^M w_{mk}^2\,|\widetilde{g}_{mk}|^2}{\sum\limits_{m = 1}^M w_{mk}^2\,( \sum\limits_{n \ne k}^K|\widetilde{g}_{mn}|^2+\sum\limits_{p=1}^K\sum\limits_{q \ne p}^K|\widetilde{g}_{mq}|^2 +\psi_k  )   }.
\end{split}
\end{eqnarray}

 For signal detection, SIC is exploited and we assume that the effective channels are arranged in an ascending order as follow \cite{ref07}:
\begin{eqnarray}
\label{5}
\sum\limits_{m = 1}^M {{{\left| {{{\widetilde g}_{m1}}} \right|}^2} \le ... \le } \sum\limits_{m = 1}^M {{{\left| {{{\widetilde g}_{mk}}} \right|}^2} \le ... \le } \sum\limits_{m = 1}^M {{{\left| {{{\widetilde g}_{mK}}} \right|}^2}},
\end{eqnarray}
and the SINR of the k-th UE can be modified as:
\begin{eqnarray}
\label{5}
\begin{split} 
{\gamma_k} =\dfrac{\sum\limits_{m = 1}^M w_{mk}^2\,|\widetilde{g}_{mk}|^2}{\sum\limits_{m = 1}^M w_{mk}^2\,( \sum\limits_{n=1}^{k-1}|\widetilde{g}_{mn}|^2+\sum\limits_{p=1}^K\sum\limits_{q \ne p}^K|\widetilde{g}_{mq}|^2 +\psi_k  )   },
\end{split}
\end{eqnarray}
and it is obvious that SIC is effective to raise the SINR with the reduction of inter-user interference.

\subsection{Problem Formulation}
We define the EE as the ratio between the normalized transmission rate (bps/Hz) and the total energy consumption (Joule).
When it comes to the total energy consumption in the entire cell-free network, it’s mainly made up of two components, the total signal transmission energy consumption (${P_K}$), the hardware energy consumption of APs (${P_{\rm{AP}}}$) and UEs (${P_{\rm{UE}}}$), respectively. As a consequence, the total energy consumption can be given by:
\begin{eqnarray}
{P_{\rm{total}}(t)} = {P_K(t)} + K {P_{\rm{UE}}} + M {P_{\rm{AP}}},
\end{eqnarray}
where ${P_K(t)} = \sum\limits_{m = 1}^M\sum\limits_{k = 1}^K w_{mk}^2(t) \tau _l\delta _l{p_u}$, and  ${P_{\rm{AP}}}$, ${P_{\rm{UE}}}$ are regarded as constant.
As the expression of SINR is given by (11), the $k$-th UE's normalized transmission rate can be derived by Shannon formula as follow:
\begin{eqnarray}
\label{5}
{R_k}(t) = {\log _{\rm{2}}}(1 + {\gamma _k}(t)).
\end{eqnarray}

The long-term EE is measured as the performance metric in the cell-free network and our goal is to find the optimal beamforming design for maximizing long-term EE. Based on the aforementioned discussion, the beamforming optimization problem to maximize the long-term EE ${\overline \eta }_{EE}$ can be formulated as follow:
\begin{subequations}\label{6}
	\begin{eqnarray}
	\label{6a}
	&\underset{{{{{\bm{W}}}_{mk}}}}{\rm max}\ &{1 \over T}\sum\limits_{t = 1}^T { {{{\sum\limits_{k = 1}^K {{{\log }_{\rm{2}}}(1 + {\gamma _k}(t))} } \over {P_K(t)+ K{P_{\rm{UE}}} + M{P_{\rm{AP}}}} }}  } \\
&\operatorname{s.t.}
&\sum\limits_{m=1}^M  w_{ml}^2(t)(|{\widetilde g}_{ml}|^2 - \sum\limits_{n=1}^{l-1}|{\widetilde g}_{mn}|^2) \geq P_s, \\
&&\sum\limits_{m = 1}^M {{w_{mk}}(t)}  = 1,\\
&&P_K(t)\le P_{\max }, \\
&&(\forall l=2,...,K~{\rm{and}}~\forall k=1,...,K) \notag
\end{eqnarray}
\end{subequations}
where ${{\bm{W}}_{mk}} \in {\left[ {0,1} \right]^{M \times K}}$ is the overall beamforming matrix. The constraint (14b) represents the successful SIC operation with the sensitivity $P_s$ of SIC receiver. The constraint (14c) represents that the beamforming vector of each UE is normalized. The constraint (14d) guarantees that the total signal transmission power is not more than ${P_{\max }}$,  where ${P_{\max }}$ is the maximum allowable total signal transmission power in cell-free networks. Since the actual channels are time-varying, and the instantaneous CSI is unavailable at CPU, we consider the channel statistics and replace channel parameters $|\widetilde{g}_{mi}|^2\,(i=k,n,q)$ and $\psi_k$ in (11) with ${\mathbb{E}}\{|\widetilde{g}_{mi}|^2\}$ and ${\mathbb{E}}\{\psi_k\}$.

The optimization problem of the long-term EE is closely bound up with long-term benefits, and the DRL algorithm is fit to figure it out.The DDPG algorithm, a branch of DRL to empower agents, interacts with the environment and promotes its learning ability. The most benefit of the DDPG algorithm is that it drives strong power to handle the problem of successive decision-making. In consequence, we utilize the DDPG algorithm to instruct CPU to perform beamforming design and maximize the long-term EE in the cell-free network.

\section{Beamforming Design using DDPG}
In this section, we tend to propose a solution for beamforming design with the DDPG algorithm and the complete design process is showed in Fig. \ref{2} in details. The DDPG algorithm configures a double-network architecture with the target and the online networks. Besides, the DDPG algorithm exploits experience mechanism and deep neural network to make the learning process more stable and the convergence rate faster.

In each training step, CPU performs the immediate beamforming design based on the current SINR and informs it to APs by control signals. Afterwards the environment of cell-free network will make reaction to the action with the rewards of all UEs, so that the environment will switch to a totally new state. In each training episode, the mini-batch stochastic gradient descent algorithm is used to train the parameters in the value network and the stochastic gradient ascent algorithm is exploited to update the parameters in the policy network. At the start of each episode, we intend to randomly generate an initial state denoted by ${s_0}$, and the environment will switch to the final state after the max-episode-steps, accompanied by a tuple of designed parameter transitions stored in the replay buffer $R$ and we select mini-batch $N$ transitions from replay buffer for parameters update.

\subsection{State-Action-Reward Construction}
As it's aforementioned in the system model, the AP locations are fixed and UEs are of low mobility. We assume that UEs' locations are static and we ignore UEs' mobility. Hence, we could regard UE's SINR only affected by the beamforming design, and our formulated problem can be modeled approximately as a Markov  decision process (MDP). We describe MDP elements first, namely state, action and reward function. The $state$ of the environment, $\bm{s_t}=\{s_1,s_2...s_k\}$, is the SINR of $K$ UEs. The $action$ is the beamforming matrix  ${{\bm{W}}_{mk}} \in {\left[ {0,1} \right]^{M \times K}}$. More importantly, the beneficial design of the reward function is closely associated with the long-term EE in the cell-free network. Actions will lead to a promotion or reduction with reward or penalty correspondingly. The $reward~function$ is designed as follow \cite{ref09}:

\begin{eqnarray}
\label{7}
\begin{split}
& r(t) = \Delta \eta(t) ={{\sum\limits_{k = 1}^K {{{\log }_{\rm{2}}}(1 + {\gamma _k}(t))} } \over {{P_K}(t) + K  {P_{UE}} + M  {P_{AP}}}}  \\
& \qquad\qquad\quad- {{\sum\limits_{k = 1}^K {{{\log }_{\rm{2}}}(1 + {\gamma _k}(t-1))} } \over {{P_K}(t-1) + K  {P_{UE}} + M  {P_{AP}}}},
\end{split}
\end{eqnarray}
and the form of reward function contributes to the convergence and performance of the DDPG algorithm empirically. 
\begin{figure}[h]
	\centering
	\includegraphics[width=3.4 in]{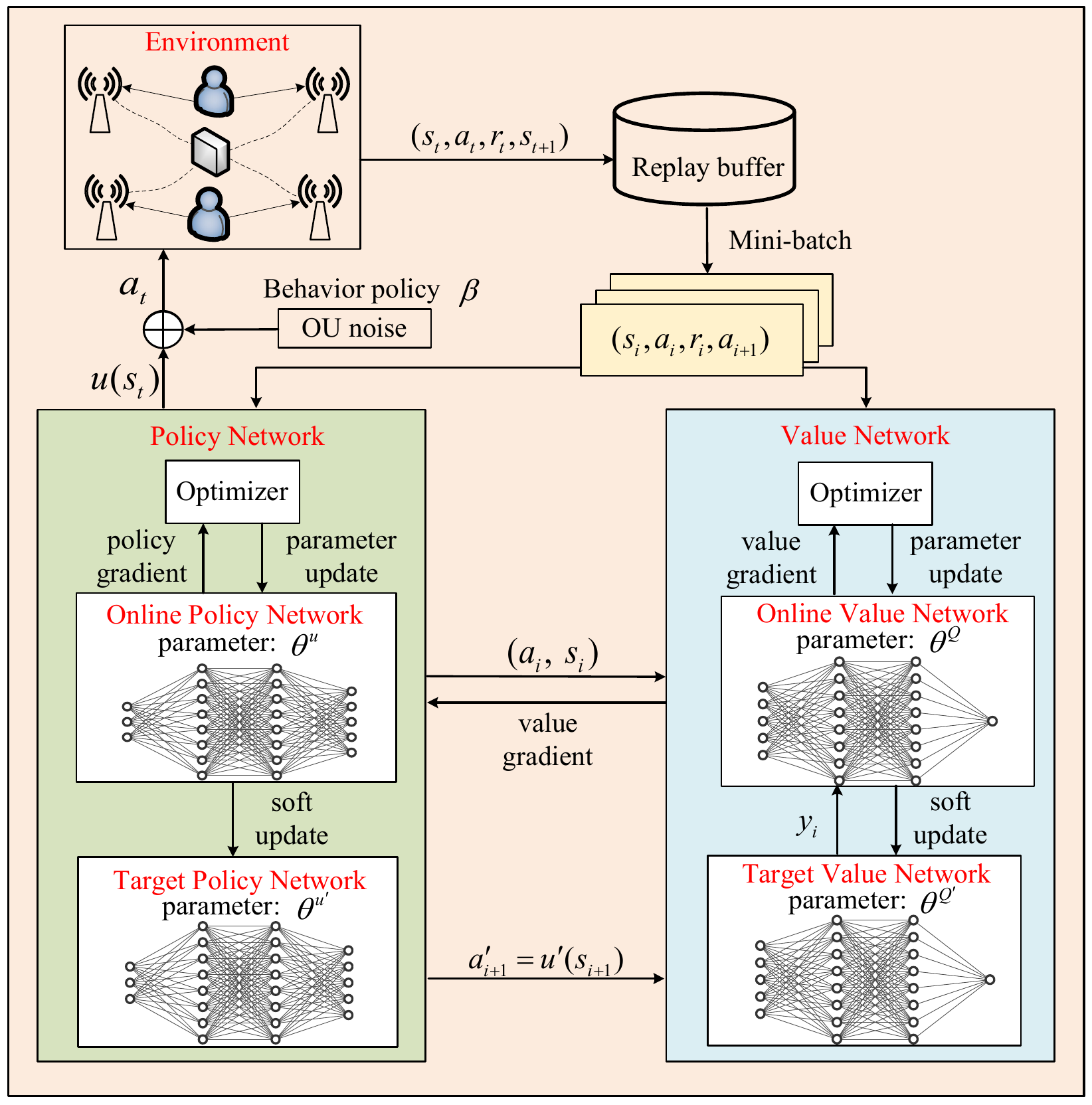}
	\caption{DDPG-based beamforming design.}
	\label{2}
\end{figure}

\subsection{DDPG-Based Beamforming Design}
The element ${w_{mk}}$ in the beamforming matrix is continuous in the range [0,1], and the DDPG algorithm provides solution to manage the problem with continuous state space and continuous action space. Thus the DDPG algorithm could be applied to search the optimal beamforming matrix ${{\bm{W}}_{mk}}$. 

The DDPG algorithm equips a double network architecture with a policy and a value network, respectively, $a = u(\bm{s}|{\theta ^u})$, $Q(\bm{s},a|{\theta ^Q})$.
In the DDPG algorithm, states map actions directly so that we're likely to ignore a probability distribution across a discrete action space. The $Q(\bm{s},a|{\theta ^Q})$ is estimated by Bellman equation as follow:
\begin{eqnarray}
\label{8}
\begin{split}
&{Q^u}({\bm{s}_t},{a_t}) = {{\mathbb{E}}_{({\bm{s}_t},{a_t},{r_t},{\bm{s}_{t + 1}}) \in R}}\{r({\bm{s}_t},{a_t}) \\
& \qquad\qquad\;+ \zeta {Q^u}({\bm{s}_{t + 1}},u({\bm{s}_{t + 1}}))\}.
\end{split}
\end{eqnarray}

Equation (16) defines the estimation of the current action value based on the current state and deterministic policy $u$, where $R$ is a set of experience and $\zeta$ is the discount factor. In order to make the DDPG algorithm more stable and efficient, it creates two neural networks for the both networks independently: the online network with parameters ${\theta ^u}$, ${\theta ^Q}$ and the target network with parameters ${\theta ^{u'}}$, ${\theta ^{Q'}}$.

The objective function is defined as the expectation of discount accumulated reward in the DDPG algorithm, which can be written as:
\begin{eqnarray}
\label{8}
{J_\beta }(u) = {{\mathbb{E}}_u}\{{r_1} + \zeta {r_2} + {\zeta ^2}{r_3} + ... + {\zeta ^{n-1}}{r_{n + 1}}\},
\end{eqnarray}
the policy $u^*$ to find optimal deterministic action is equivalent to the policy of maximizing objective function ${J_\beta }(u)$ as $u^* = {\rm{argma}}{{\rm{x}}_u}{J_\beta }(u)$ \cite{DDPG} and the gradient of the policy network is:
\begin{eqnarray}
\label{8}
{\nabla _{{\theta ^u}}}J \approx {1 \over N}\sum\limits_i {{\nabla _a}Q(\bm{s},a){|_{\bm{s} = {\bm{s}_i},a = u({\bm{s}_i})}}{\nabla _{{\theta ^u}}}u(\bm{s}){|_{\bm{s} = \bm{s}_i}}},
\end{eqnarray}
and we optimize the objective function by stochastic gradient ascent algorithm with learning rate $lr_u$. In the online value network, the gradient can be represented as:
\begin{eqnarray}
\label{8}
{\nabla _{{\theta ^Q}}} = {1 \over N}\sum\limits_i {[(y_i - Q({\bm{s}_i},{a_i})){\nabla _{{\theta ^Q}}}Q({\bm{s}_i},{a_i})]},
\end{eqnarray}
where $y_i = {r_i} + \zeta Q'({\bm{s}_{i + 1}},u'({\bm{s}_{i + 1}}|{\theta ^{u'}})|{\theta ^{Q'}})$, and we update the online value network by stochastic gradient descent with learning rate $lr_Q$. Finally, $\theta ^{u'}$ and $\theta ^{Q'}$ in the target networks are updated by Poylak averaging factor $\tau $:
\begin{subequations}\label{6}
	\begin{eqnarray}
	\label{6a}
	&{\theta ^{Q'}} = \tau {\theta ^Q} + (1 - \tau ){\theta ^{Q'}},\\
	&{\theta ^{u'}} = \tau {\theta ^u} + (1 - \tau ){\theta ^{u'}}.
	\end{eqnarray}
\end{subequations}

In conclusion, the ultimate aim of the DDPG algorithm is to maximize the objective function ${J_\beta }(u)$ in the policy network and minimize the loss of the action value Q in the value network simultaneously. \textbf{Algorithm 1} summarizes the DDPG-based beamforming design.

\begin{algorithm}[t]
	\caption{DDPG-Based Beamforming Design}
	\label{Algorithm1}
	\begin{algorithmic}[1] 
		\STATE Randomly \textbf{initialize} the value network $Q(\bm{s},a|{\theta ^Q})$ and the policy network $a = u(\bm{s}|{\theta ^u})$ with weights ${\theta ^Q}$ and ${\theta ^u}$;
		\STATE \textbf{Initialize} the target value network $Q'$ and the target policy network $u'$ with weights ${\theta ^{Q'}} = {\theta ^Q}$ and ${\theta ^{u'}} = {\theta ^u}$;
		\FOR{episode = 1 to Max-number-episodes}
		\STATE Randomly \textbf{initialize} process ${\cal N}$ for action exploration
		\STATE \textbf{Initialize} replay buffer $R$ and randomly generate ${\bm{s}_0}$;
		\FOR{t=1 to Max-episode-steps}
		\STATE CPU executes the beamforming design based on the state $\bm{s}_t$ and the policy $u$, and  ${a_t} = u({\bm{s}_t}|{\theta ^u}) + {{\cal N}_t}$;
		\STATE APs perform the action ${a_t}$ and CPU record reward ${r_t}$ and the next state ${\bm{s}_{t+1}}$;
		\STATE Store the transition $({\bm{s}_t},{a_t},{r_t},{\bm{s}_{t + 1}})$ in $R$;
		\ENDFOR
		\STATE Randomly sample mini-batch of $N$ transitions from $R$, where $N$ is the mini-batch size;
		\STATE Minimize the loss function to update the online value network:
		\begin{center}${\rm{Loss}} = {1 \over N}{\sum\nolimits_i {(y_i - Q({\bm{s}_i},{a_i}|{\theta ^Q}))} ^2}$,\end{center}
		\begin{center}$y_i = {r_i} + \zeta Q'({s_{i + 1}},u'({\bm{s}_{i + 1}};{\theta ^{u'}})|{\theta ^{Q'}})$,\end{center}
		\begin{center}${\nabla _{{\theta ^Q}}} = {1 \over N}\sum\nolimits_i {[(y_i - Q({\bm{s}_i},{a_i})){\nabla _{{\theta ^Q}}}Q({\bm{s}_i},{a_i})]}$,\end{center}     
		and the update formula of ${\theta ^Q}$ can be expressed as:
		\begin{center}${\theta ^Q} = {\theta ^Q} - lr_Q \cdot {\nabla _{{\theta ^Q}}}$;\end{center}
		\STATE Update the online policy network by sampled stochastic policy gradient ascent as:
		\begin{center}${\nabla _{{\theta ^u}}}J \approx {1 \over N}\sum\limits_i {{\nabla _a}Q(\bm{s},a){|_{\bm{s} = {\bm{s}_i},a = u({\bm{s}_i})}}{\nabla _{{\theta ^u}}}u(\bm{s}){|_{\bm{s} = \bm{s}_i}}}$,\end{center}
		and the update formula of ${\theta ^u}$ can be expressed as:
		\begin{center}${\theta ^u} = {\theta ^u} + lr_u \cdot {\nabla _{{\theta ^u}}}$;\end{center}
		
		\STATE Soft update the target value network and the target policy network by Poylak averaging factor $\tau$ as follow:\\
		\begin{center}${\theta ^{Q'}} = \tau {\theta ^Q} + (1 - \tau ){\theta ^{Q'}}$,\end{center}
		\begin{center}${\theta ^{u'}} = \tau {\theta ^u} + (1 - \tau ){\theta ^{u'}}$.\end{center}		
		\ENDFOR		 	
	\end{algorithmic}  
\end{algorithm}

\subsection{Complexity Analysis} 
In this section, we’re about to discuss the complexity analysis of the DDPG algorithm. Under the assumption that $M$ APs and $K$ UEs are considered in the cell-free network and the overall beamforming matrix, ${{\bm{W}}_{mk}} \in {\left[ {0,1} \right]^{M \times K}}$, demands optimization. When we set a certain step size $\Delta $ and the conventional approaches for the beamforming design follow the step, thus the complexity of the conventional approaches, $O\left( {{{({1 \over \Delta })}^{M + K}}} \right)$, is exponential. The complexity analysis of the DDPG algorithm depends on two aspects: inference floating operations per second (FLOPS) and the convergence rate. The number of FLOPS during the inference is mainly determined by the structure of the policy network and the value network. When we set $\left| S \right|,\left| A \right|,\left| {{H_i}} \right|$ as the numbers of elements in state, action and the n-th hidden layers in policy and value network, the inference FLOPS in the policy network and the value network can be computed as:
\begin{subequations}\label{6}
	\begin{eqnarray}
	\label{6a}
	&{\rm{FLPOS}_{u}} = | S | | {{H_1}}|  + \sum\limits_I {| {H_{i - 1}} H_i |}+| A |  | {{H_I}} |,\\
	&{\rm{FLPOS}_{Q}} = {| S  +  A |} | {H_1} | + \sum\limits_I {| {H_{i - 1}}  {H_i} |}+|H_I|,
	\end{eqnarray}
\end{subequations}
where $I$ represents the number of hidden layers in each neural network. As a consequence, the number of FLOPS during the inference results in low complexity because of its scalar multiplication in the DDPG algorithm. Correspondingly, the convergence rate is faster than the conventional approaches and it is shown in Fig. 3. In general, the DDPG algorithm is a more superior scheme to handle our formulated problem.
\section{Simulation Results}
In numerical analysis, we evaluate the EE with our proposed DDPG-based beamforming design in the cell-free network. We consider a possible cell-free network size with $M=10$, $K=6$ where APs and UEs are uniformly located in a range of radius $r=20$ meters. The hardware power consumption of APs and UEs are both 20 dBm, pilot length ${\tau _l}=6$ samples, pilot transmission power per symbol ${\delta _l}=20$ dBm, uplink transmission power $p_u=16$ dBm, SIC sensitivity $P_s=1$ dBm \cite{ref07} and noise power ${\sigma ^{\rm{2}}}=-80$ dBm. We train our proposed model with the tool of Python and Pytorch 1.4.0, and the number of training episode is 1000 with 200 steps in each episode. The networks both have two fully-connected hidden layers with size 256$\times$128. The activation function is leaky $Relu$ function in each layer, and the output layer is $softmax$ function in the policy network and none in the value network. The loss function is Mean Squared Error in the value network and the gradient of Q-value is used in the policy network update. $Adam$ is employed as the optimizer of the networks.
The hyper-parameters of the DDPG algorithm are set as follows: the discount factor $\zeta  = 0.7$, the learning rate $lr_u = 0.01$ and $lr_Q = 0.02$, the Poylak averaging factor $\tau  = 0.006$, the size of mini-batch $N=32$, and the size of replay buffer $R = {10^5}$.

\begin{figure*}[t]
	\centering
	\begin{minipage}[t]{0.22 \linewidth}
		\centering
		\includegraphics[width=1.8 in]{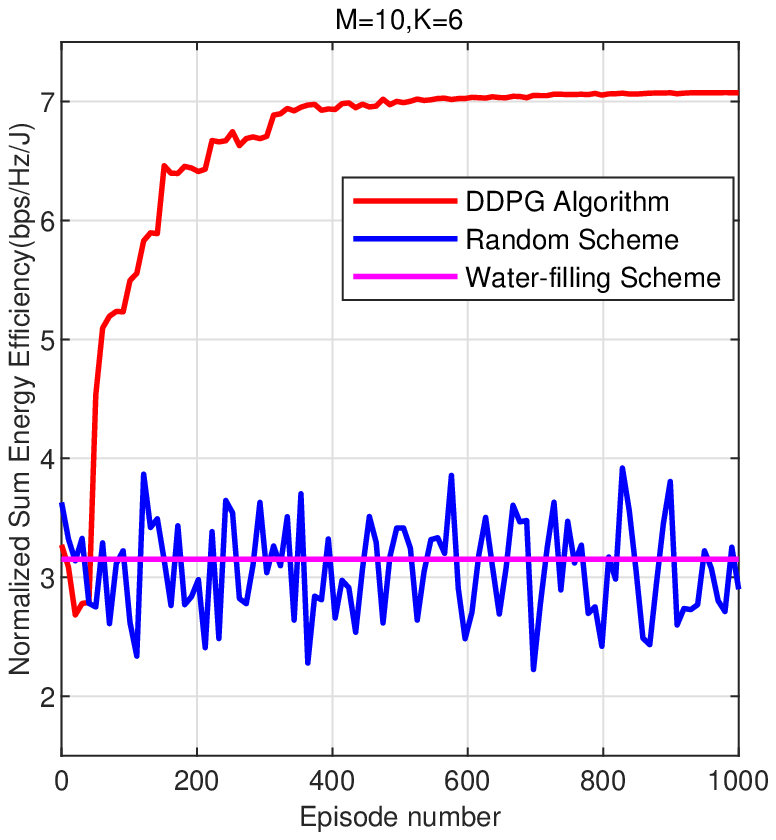}
		\caption{Convergence of EE.}
		\label{EE6}
	\end{minipage}
	\hspace{0.3cm}
	\begin{minipage}[t]{0.22 \linewidth}
		\centering
		\includegraphics[width=1.8 in]{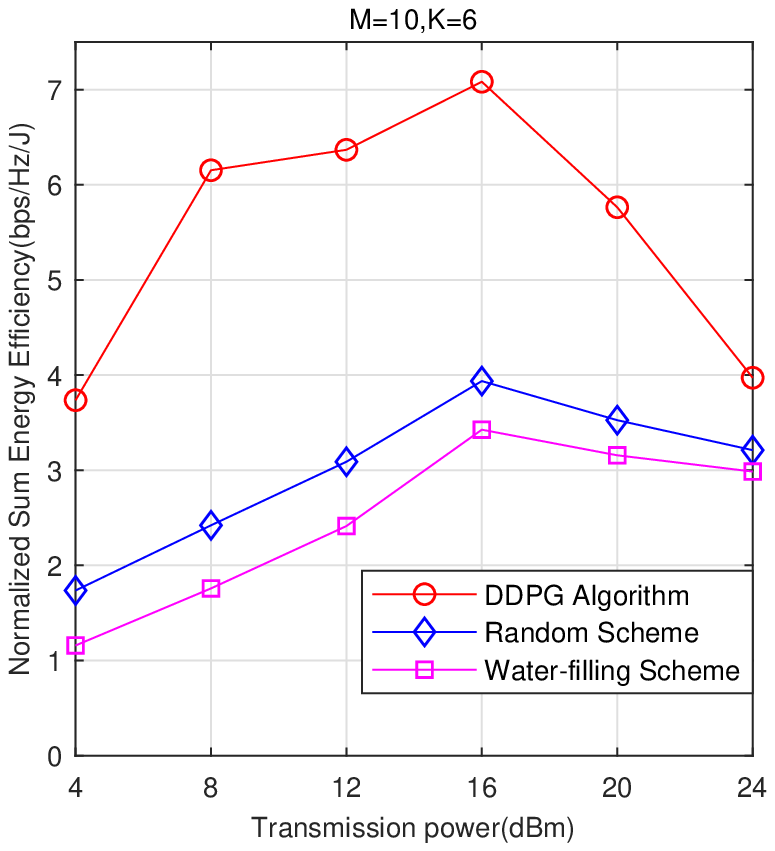}
		\caption{EE vs. transmission power.}
		\label{PM}
	\end{minipage}
	\hspace{0.3cm}
	\begin{minipage}[t]{0.22 \linewidth}
		\centering
		\includegraphics[width=1.8 in]{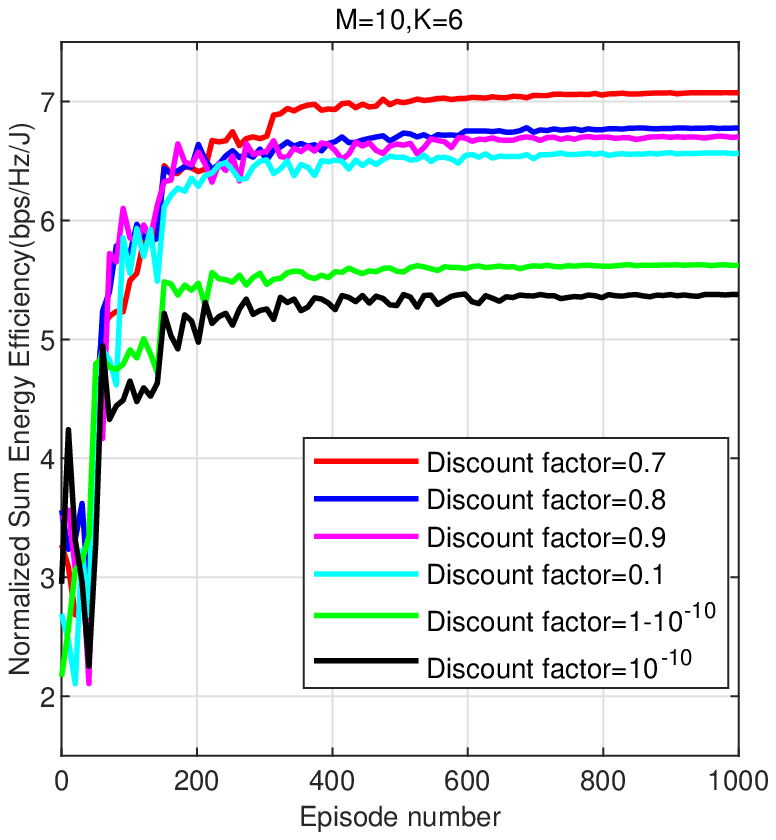}
		\caption{EE vs. discount factor.}
		\label{GAMA}
	\end{minipage}
	\hspace{0.3cm}
	\begin{minipage}[t]{0.22 \linewidth}
		\centering
		\includegraphics[width=1.8 in]{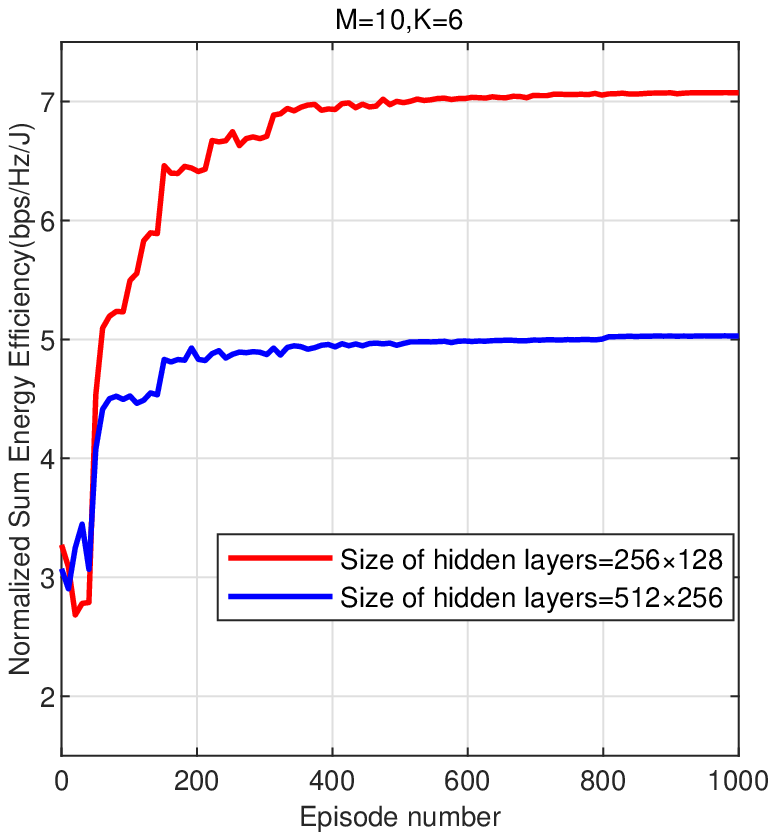}
		\caption{EE vs. hidden layers size.}
		\label{SIZE}
	\end{minipage}
\end{figure*} 

Next, we verify the effectiveness of the DDPG algorithm and the EE performance of our proposed DDPG-based beamforming design in the cell-free network. Furthermore, we take the water-filling scheme (larger $g_{mk}$ determines larger ${w_{mk}}$) and the random scheme as the benchmarks. We intend to finish the simulation work in terms of three aspects:

\subsubsection{Convergence of the DDPG-Based beamforming design}
Fig. 3 illustrates the convergence of the DDPG-based beamforming design in the training episodes. We may draw a conclusion that the DDPG algorithm is capable of converging over the 1000 episodes, while the EE performance of the other two methods remain poor over 1000 episodes. In addition, the EE performance of the DDPG-based beamforming design achieves 90\% of the best performance over about 180 episodes and the EE performance gap expands since about 50 episodes.

\subsubsection{Energy efficiency versus transmission power}

Fig. 4 characterizes the EE versus uplink signal transmission power $p_u$. In the simulation, we tend to observe the variation of EE performance when $p_u$ ranges from 4~dBm to 24~dBm. From Fig. 4, EE rises as $p_u$ increases from 4~dBm to 16~dBm and declines from 16~dBm to 24~dBm. It is because that the EE raises as the $p_u$ satisfies the demand of signal transmission, while when the $p_u$ is high enough, the redundant energy consumption causes EE performance decline. 

\subsubsection{The influence of the hyper-parameters }
Fig. 5. illustrates that $\zeta  = 0.7$ hits the optimal EE performance and it degrades in an acceptable range when  $\zeta  = 0.{\rm{1}},{\rm{0}}{\rm{.8}},{\rm{0}}{\rm{.9}}$. However, the extreme $\zeta$  will lead to egregious EE performance: when $\zeta  = {\rm{1}}{{\rm{0}}^{{\rm{ - 10}}}}$, the Bellman equation closely associates with the instantaneous reward; when $\zeta  = 1 - {\rm{1}}{{\rm{0}}^{{\rm{ - 10}}}}$, the Bellman equation will stand for the one-day’s sum reward to a large extent which leads to a poor EE performance. 
Fig. 6 shows that the hidden layers with the neuron size of 256$\times$128 results in a better EE performance than the neuron size of 512$\times 256$. It is because that if we set redundant neurons in hidden layers, it may occur overfitting which causes stuck in local minima instead of global optimal.

\section{Conclusion}
This paper investigated the DRL for energy-efficient beamforming design in cell-free networks. Based on MMSE channel estimation and SIC signal detection technologies, the closed-form of SINR per user and long-term EE function were derived. The DDPG algorithm was exploited to perform centralized beamforming design for the long-time EE maximum problem with continuous state and action space. It was demonstrated that the DDPG-based algorithm was convergent and reduced the exponential computational complexity to polynomial level. The simulation results indicated that the DDPG-based beamforming design outperformed benchmarks in terms of EE under different network setups. Moreover, appropriate discount factor and hidden layers size could lead to preferable performance. 


\bibliographystyle{IEEEtran}
\bibliography{IEEEabrv,reflwl} 

\end{document}